# Numerical modeling of rocking of shallow foundations subjected to slow cyclic loading with consideration of soil-structure interaction

S. Mohsen Haeri and Aria Fathi
Civil Engineering Department, Sharif University of Technology, Tehran, Iran.

**ABSTRACT**

Strong Vibration of buildings during seismic loading may result in an uplift or partial separation of the foundation from the underneath soil. To date, various researches have indicated that Soil-Structure Interaction (SSI) has many favorable features including a probable increase in natural period of soil-structure system and also a decrease in shear base demand in structures. Furthermore, Rocking is one of the most important factors in describing the rotational behavior of a structure built on a shallow foundation especially on a soft soil which can affect the dynamic behavior of the structure noticeably. To study the effects of Rocking of shallow foundations subjected to slow cyclic loading with consideration of soil-structure interaction, a Finite Element Method (FEM) using ABAQUS software has been deployed to simulate the rocking motion of shallow foundations. For a more efficient simulation of the soil, both linear and non-linear elasto-plastic behavior of the soil have been taken into account in the analysis using the sub-routine coded in FORTRAN. The results achieved with observations notably show that allowing the foundation to rock may result in stiffness degradation of the soil-structure system and an increasing energy dissipation of soil-structure, especially in high rise structures. Additionally, results describe that deploying the linear elastic–perfect plastic approach may result in higher uplift of the foundation in comparison to that using a non-linear elasto-plastic approach, particularly in structures with lower heights.

**INTRODUCTION**

In these recent years, many attempts have been carried out to predict the behavior of soil during dynamic loading such as earthquake. These predictions lead to more efficient design of structures which located on soil. One of the most important aspects of seismic design which attracts a lot of attentions among engineers, is recognition of the Soil-Structure Interaction (SSI), and significant effects of these corresponding systems on seismic behavior of each other. To study the SSI having a clear conception of the rocking mode should be necessary in order to scrutinize the dynamic behavior of shallow foundations located on flexible soils, therefore, several researches have been taken into account in this regard. Primary researches on rocking have been inaugurated since early 1960s when Housner [1] - with considering his observations over severe earthquakes- investigated in rigid blocks subjected to horizontal loading so as to study the separation of them from underneath surface. His obtained results show that structures which are able to rock were less damaged in comparison to apparently more stable and state-of-the-art ones [1].

On the one hand Conventional engineered-designs methods emphasize on preventing the overturning of structures due to rotation during strong vibrations, on the other hand, recent studies indicate that allowing the structures to rock may result in energy dissipation by the soil-structure system, deploying the SSI effects [2],[3].

In general, having many beneficial effects during seismic loading such as probable increase in natural period of soil-structure system and also a decrease in shear base demand in structures, the non-linear behavior of the soil plays a key role to dissipate the energy during strong vibrations; however, permanent displacements of foundation are the most noticeable disadvantage of nonlinearity which leads to inevitable damages [2], [3].

Importance of the rocking mode and nonlinearity of the soil are well-known after all these researches over load-displacement behavior of shallow foundations which occasionally result in new seismic design philosophies. New approaches signify an intentional under-designing the foundation to take advantage of the development of plastic hinging at the foundation instead of the superstructure (e.g. I. Anastasopoulos et al. 2009 [4]) or to attribute both superstructure and foundation jointly as an energy-dissipation mechanisms -Balanced Design (BD) system- as an alternative to Structural Hinging Dominated (SHD) and Foundation Rocking Dominated (FRD) systems (Weian Liu et al. 2013 [5]). Accordingly, these novel methods could prevent catastrophic collapse of the structures during strong vibrations.

In this paper, rotational behavior of 10, 20 and 30 story structures subjected to slow lateral cyclic loading, deploying SSI, have been taken into consideration. To study the rocking effects of shallow foundations, the Finite Element Method (FEM) has been used. Additionally, two linear and non-linear elastic-perfect plastic approaches have been taken into account for a more efficient simulation of the soil by taking advantage of two sub-routines coded in FORTRAN. Three series of analyses including nine numerical models have been carried out using ABAQUS program in order to assess energy dissipation of the soil-structure system, critical contact area, permanent settlement of shallow foundation and uplift, are presented in the following sections.

## MODEL ANALYSIS PROGRAM

To study the rocking behavior of shallow foundations, nine numerical Models including soil, structure and foundation systems were simulated. These models categorized in three main series to investigate the linear and non-linear elasto-plastic behavior of the granular soil, and to consider the soil stiffness variation effects upon rotational behavior of soil-structure system in the 30 story structure in particular.

In this research to study and compare the high rise structures three structures with different heights were investigated. These analyzed structures have a lot characteristics in common, such as length (taken to be 20 m) and width dimensions, elements type, and material properties; yet height dimension is different among them. Building Model as an integrated rigid system consists of a foundation and the structure located on it. The mass of structure in conjunction with its foundation evaluated based on the mass of conventional structures. The used structures Parameters are shown in Table 1.

Table 1 Structure parameters

| Structure Properties | 10 Story | 20 Story | 30 Story |
|---|---|---|---|
| Height (m) | 30 | 60 | 90 |
| Weight (KN) | 16620 | 39240 | 58860 |
| Number of Elements | 6850 | 11800 | 13800 |
| Elastic Modulus (GPa) | 25 | 25 | 25 |
| Poison Ratio | 0.35 | 0.35 | 0.35 |

The Mohr-Coulomb model has deployed in all cases for the soil underneath the Foundation. Although, there is a general consensus among researchers that shear modulus of soils, particularly the granular soils, vary with both shear strain levels and confining pressure (Kramer 1996 [6]). However, in majority of dynamic numerical analyses, this intrinsic feature of soil is considered at static shear strain level. Therefore, for a more realistic results, the nonlinearity and inhomogeneity of the soil would be taken into consideration by using the sub-routine coded in FORTRAN. The maximum shear modulus of the soil ($G_{max}$), obtained from Eq. (1), proposed by Seed and Idriss 1970 [7].

$$G_{max(KN/m^2)} = 218.82 \cdot K_{2(max)} + (\sigma')^{0.5} \qquad (1)$$

$$\sigma' = \frac{\sigma_1' + \sigma_2' + \sigma_3'}{3} \qquad (2)$$

Where $\sigma'$ is the effective confining stress, obtained from Eq. (2); $\sigma_1', \sigma_2'$ and $\sigma_3'$ are the maximum, intermediate and minimum effective principal stresses, respectively; and the magnitudes of $K_{2(max)}$ vary from 35 to 70 for sandy soils (taken to be 52 for all cases which refers to fine granular soils with relatively high relative density).

In an attempt to calculate shear modulus at corresponding shear strain ($\gamma'(\%)$) and its variation with confining pressure, studies by Seed et al. (1986) was deployed [8]. Thus, with all this taken into account, in the first series of analyzed cases the nonlinear elastic-perfect plastic behavior are considered for the soil.

Also, in the second series of analyses, the linear elasto-plastic behavior is allocated to soil in order to show response of the soil act as a linear-elastic before the mobilized shear stress reaches to the Mohr-Coulomb Failure Envelope, and perfectly plastic when the soil yields by reaching the shear stress to the shear strength. Supposing the shear modulus was used at the static shear strain ($\gamma'(\%) = 10^{-2}$), whereas, the shear modulus is varied by the confining pressure.

In dynamic analysis, damping plays a key role by dissipating a part of elastic energy. With this in mind, Rayleigh viscous damping on account of its simple formulation could be effective to express the elastic energy dissipation. That is why in this study, following formulation proposed by Rayleigh and Lindsay (1945) has been used.

$$C = \alpha \cdot [M] + \beta \cdot [K] \qquad (3)$$

In which [C] = damping matrix of physical system; [M] = mass matrix of physical system; [K] = stiffness matrix of the system; $\alpha$ and $\beta$ are Rayleigh damping coefficients which can be computed by two significant natural modes using standard form of Eq. (3), is mentioned in Eq. (4).

$$2\zeta_i \omega_i = \alpha + \beta \omega_i^2 \qquad (4)$$

Where $\zeta$ is damping ratio (taken to be 8% for all the soil systems and 5% for all the structure systems).

One of the most remarkable measures so as to obtain Rayleigh damping coefficients, is choosing the most efficient natural frequency modes. The first mode as well as the first odd mode which its frequency is higher than the loading frequency should be considered for solving the Eq. (4) which is suggested by Park an Hashash (2004)[9].

**Lateral Cyclic Displacements on Structures**

Figure 1 shows a schematic view of soil-structure system in conjunction with the external loads applied to the structure and the forces which are translated to the base center point of the soil-structure interface.

Based upon the Fig.1, the external loads are obtained by:

$$P_h = P_{act} \tag{5a}$$

$$M = P_h.h_{cg} + M_s.g.h_{cg}.\sin(\theta) \tag{5b}$$

Where $P_{act}$ is reaction force which is obtained during the applied lateral cyclic displacement; $M_s$ is mass of the structure; $M$ is moment acts at the base center point; $h_s$ is monolithic height of structure and foundation; $h_{cg}$ is center of gravity height of the structure-foundation system; g is gravitational acceleration; and $\theta$ is rotation of the structure.

Figure 2 shows the sinusoidal slow lateral cyclic displacements were taken into consideration in analyzed structures. To assess the stiffness variation of the soil-structure systems at their interface, all the time history-displacements apply constant rotation to each of the structures during loading. The slow cyclic time history-displacements include three clusters with different amplitudes, that each of the clusters have three cycles as shown in Fig. 2.

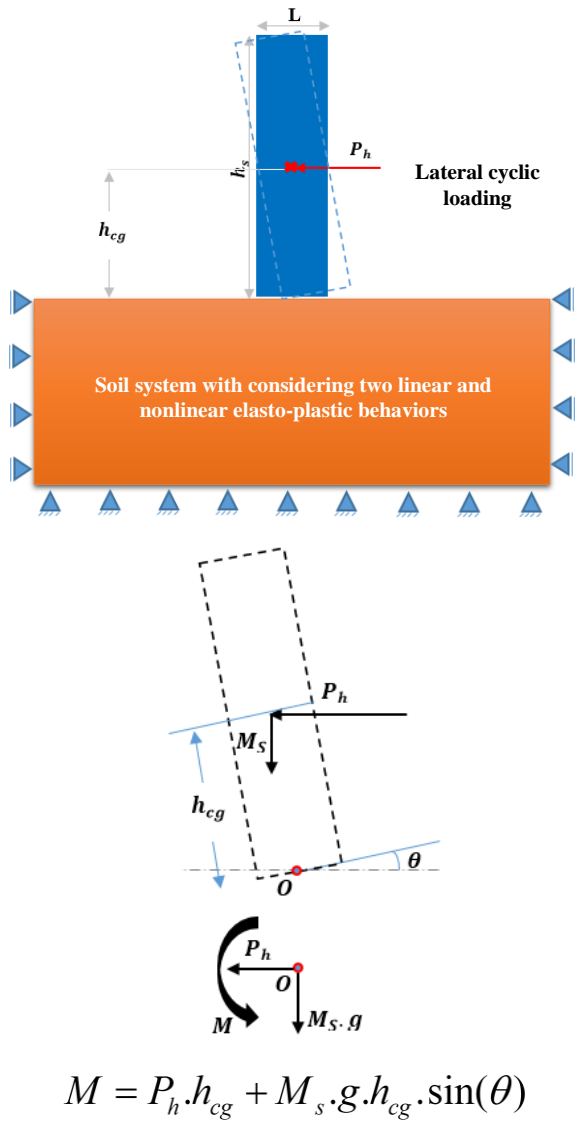

Fig. 1 Schematic model of soil-structure system as well as the external loads applied on the structure.

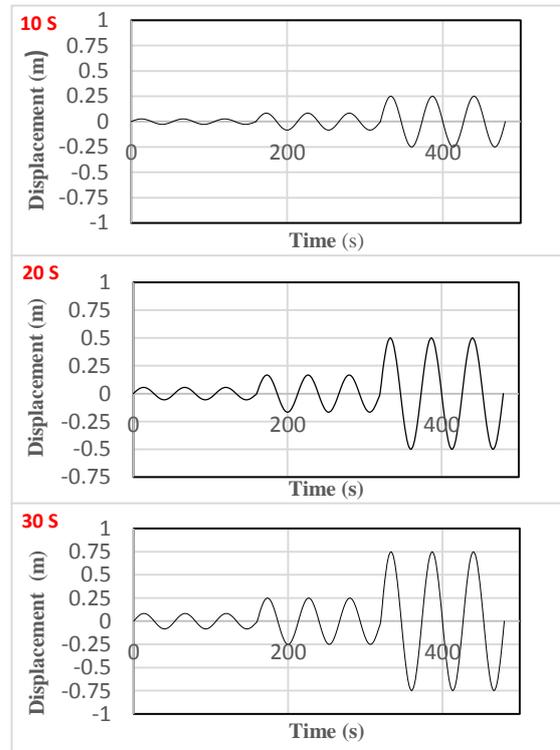

Fig.2 Slow lateral time history-displacements for 10, 20, and 30 story analyzed structures.
**Notice: S stands for story

**Verification**

The finite element model of rocking behavior of shallow foundations, deploying SSI effects has been verified using S21 test results which is related to second series of centrifuge tests (Krr02), which had been provided by K. R. Rosebrook & B. L. Kutter [10]. In S21 test, a double-wall structure configuration which includes two Aluminum shear walls with their parallel aluminum strip footings, had been considered. The Nevada sand having a relative density of 60 percent was taken into account in 20 centimeter height container. The soil characteristics as well as mechanical properties of the shear wall structures was described in Ref. [10]. The centrifuge acceleration was 20 g, is used to convert data to prototype scale; the soil-structure system for the S21 test is shown in prototype scale in Fig. 3. By the way, the nonlinear elastic-perfect plastic behavior is considered for the soil system in numerical simulation.

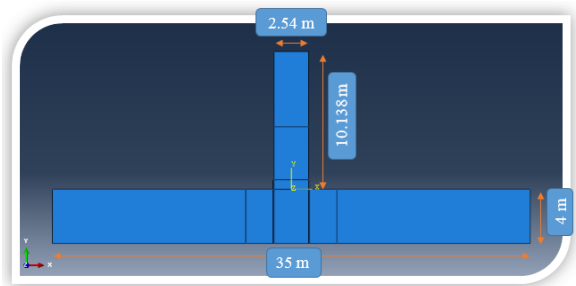

Fig. 3  The prototype scale of the soil-structure system for the S21 test.

The applied sinusoidal slow cyclic time history-displacement for this centrifuge test consists of three cycles as shown in Fig. 4

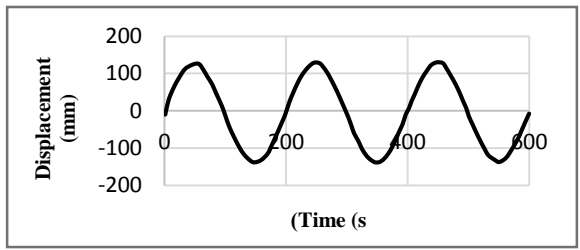

Fig. 4  Applied time history-displacement for the S21 centrifuge test [10].

*Comparison of numerical model and centrifuge test*

Figure 5 shows the rotation-moment relationship of soil-structure system at the base center point footing, computed by proposed method, and is juxtaposed with the centrifuge test results proposed by K. R. Rosebrook & B. L. Kutter [10]. The results of the numerical model are in good agreement with the experimental model, as could be seen in Fig. 5.

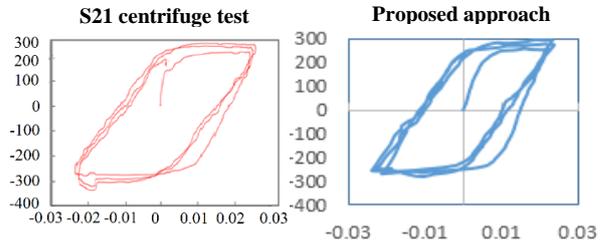

Fig. 5  Comparison of rotation-moment relationship for the aluminum shear wall structure with the obtained results based on centrifuge test by Rosebrook & Kutter [10].

**ANALYSIS RESULTS**

**Static Analysis Results**

The initial settlements of soil underneath the base center point of foundation for 10, 20, and 30 story structures are computed, considering both linear and nonlinear elastic-perfect plastic behaviors for the granular soil; and are compared with the approximate initial settlement method proposed by Mayne and Poulos (1999)[11]. The results show that the initial settlement in linear approach is closer to the approximate method results in comparison to the nonlinear approach which are shown in Table 2.

Table 2 Initial settlement of soil computed by three different approaches.

| Structure Properties | 10 Story | 20 Story | 30 Story |
|---|---|---|---|
| Approximate method | 9.2 | 18.4 | 27.6 |
| Linear approach | 7.8 | 15.0 | 23.4 |
| Nonlinear approach | 57.5 | 72.0 | 82.0 |

The difference between the initial settlement results in nonlinear elasto-plastic approach with those obtained based on two other methods which are observed in Table 2, is on account of yielding of the soil when shear modulus decrease at corresponding high shear strain levels.

**Dynamic Analysis Results and Rocking Behavior**

*Energy dissipation and rotational stiffness degradation*

It is crystal clear that part of the input energy is dissipated by the rocking motion of shallow foundation in all analysis, and are observed in rotation-moment diagrams as shown in Fig. 6. In this regard, the hysteresis loops are thicker when the nonlinear elasto-plastic behavior is considered for the soil beneath the foundation. Indeed, yielding of the soil at sides of the foundation during the rocking mode owing to increasing the shear stress and decreasing the shear modulus in these areas, eventuate in more energy dissipation through the nonlinear approach. The rotational stiffness degradation in nonlinear approach is slightly more, compare with the linear one, the stiffness degradation as well as softening of the soil system are associated with increasing the rotational amplitudes (Fig. 6).

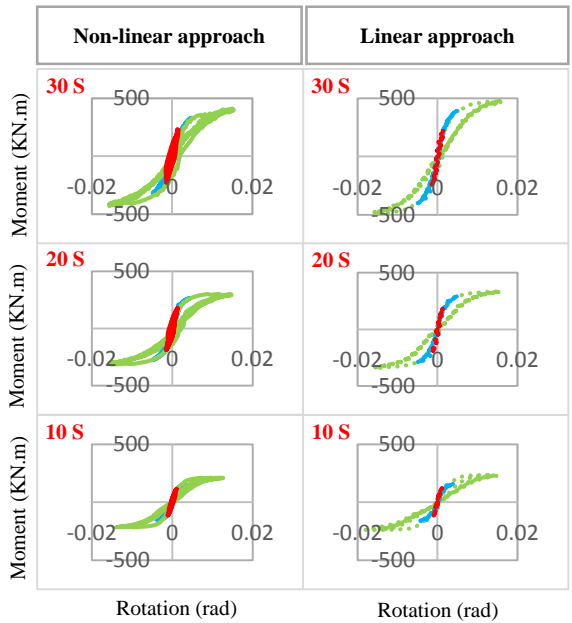

Fig. 6  Rotation-moment diagrams for the 10, 20, and 30 story structures with consideration of two linear and non-linear elasto-plastic approaches for the soil underneath the foundation.

Additionally, Fig. 6 shows that energy dissipation of soil-structure system during rocking could be enhanced not only by increasing the rotational amplitudes but also by increasing the height of structures.

The plotted soil settlement-time history for the base center point of soil-foundation's interface has been put on the foundation displacement-time history for the same point, and are shown in Fig. 7. It is evident from these two plots that a great amount of input energy is dissipated through the nonlinear elaso-plastic behavior for the soil and especially in the tall, slender structure (30 story structure). In fact, the compatibility of soil settlement and foundation displacement plots show that the input energy can be dissipated by yielding of the soil, which result in more permanent settlement and lower reflection of foundation which is observed by uplift of the foundation in comparison to stiffer behavior.

Moreover, in the first two clusters of loading, both curves are more compatible with each other than the third cluster which its amplitude is greater.

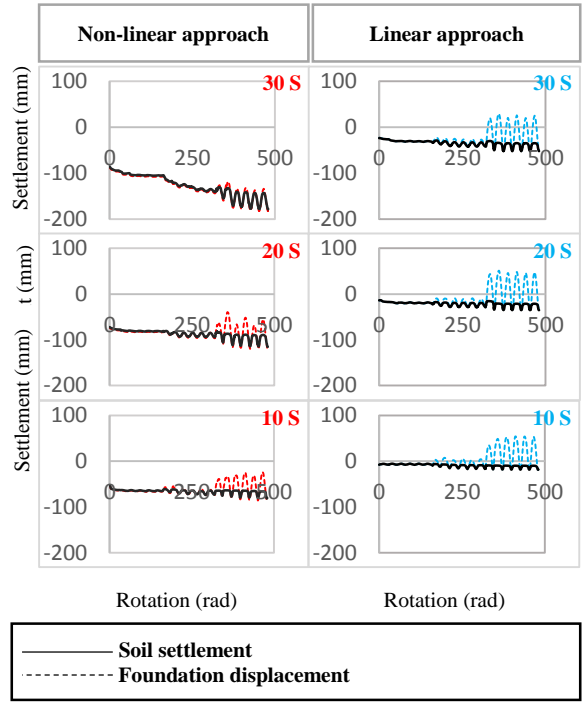

Fig. 7  Comparison of settlement and displacement-time history plots for the base center point of soil-foundation interface through the nonlinear and linear approaches.

*Mobilized Moment and foundation contact area*

In each cycle of loading, the foundation contact area is varied by the applied rotation, therefore, mobilized moment, energy dissipation, and uplift of foundation during rocking could have a close relation with ratio of foundation contact area to foundation area which is defined by following equation.

$$\zeta = \frac{A_C}{A} = \frac{Foundation\ Contact\ Area}{Foundation\ Area} \qquad (6)$$

Critical amount of $\zeta$ as well as its corresponding mobilized moment of soil-structure system are computed at maximum rotational amplitudes of each cluster of loading, are shown in table 3. In this regard, maximum mobilized moment in each cluster of loading is obtained, where the foundation has the least contact area with the soil underneath. Based on obtained results for $\zeta$, can be understood that the soil-structure system with considering the nonlinear behavior, will be reached to ultimate moment capacity at higher rotational amplitudes, and shows that the stability of structures might be enhanced against overturning compare with the linear approach.

Table 3  The ratio of foundation contact area to foundation area as well as the maximum mobilized moment.

|  |  |  | $\zeta = \dfrac{A_C}{A}$ | | | Maximum Mobilized Moment (MN.m) | | |
|---|---|---|---|---|---|---|---|---|
|  |  |  | 10 S | 20 S | 30 S | 10 S | 20 S | 30 |
| **Non-linear approach** | **Rotation (rad)** | 0.0015 | 0.83 | 0.87 | 0.96 | 115.76 | 178.08 | 229.09 |
|  |  | 0.005 | 0.45 | 0.48 | 0.66 | 152.12 | 265.64 | 331.80 |
|  |  | **0.015** | **0.33** | **0.42** | **0.45** | **212.00** | **301.17** | **411.41** |
|  |  | End of Loading | 0.63 | 0.87 | 0.92 | 0 | 0 | 0 |
| **Linear approach** | **Rotation (rad)** | 0.0015 | 0.71 | 0.80 | 0.84 | 123.78 | 188.24 | 240.88 |
|  |  | 0.005 | 0.37 | 0.39 | 0.47 | 160.11 | 289.39 | 396.65 |
|  |  | **0.015** | **0.22** | **0.23** | **0.26** | **236.06** | **336.07** | **483.11** |
|  |  | End of Loading | 0.47 | 0.53 | 0.58 | 0 | 0 | 0 |

**CONCLUSION**

The obtained results from the slender, high rise structures, which are allowed to rock show that the stiffness degradation of the soil-structure system decreases as well as increasing the energy dissipation. Although, the nonlinear elastic-perfect plastic approach result in dissipating the major amount of the input energy in comparison to the linear elastic-perfect plastic behavior considered for the soil, yet it causes the noticeable permanent settlement which might have destructive effects on the structures. Additionally, deploying the linear approach through the analysis eventuate in the less contact area during rocking, and it can make structures less stable against overturning. To sum up, with all this taken into account, rocking and nonlinearity of the soil have many beneficial effects for the soil-structure system during strong vibrations that should be more noticed in future engineered designs.